\documentclass[fleqn,twoside]{article}
\usepackage[headings]{espcrc2}

\readRCS
$Id: espcrc2.tex,v 1.2 2004/02/24 11:22:11 spepping Exp $
\usepackage{graphicx}
\usepackage[figuresright]{rotating}

\newcommand{\AmS}{{\protect\the\textfont2
  A\kern-.1667em\lower.5ex\hbox{M}\kern-.125emS}}

\newcommand{\be}{\begin{equation}}
\newcommand{\ee}{\end{equation}}
\newcommand{\bea}{\begin{eqnarray}}
\newcommand{\eea}{\end{eqnarray}}
\newcommand{\beq}{\begin{equation}}
\newcommand{\eeq}{\end{equation}}

\newcommand{\mi}{microcalcification }
\newcommand{\mis}{microcalcifications }

\title{A scalable Computer-Aided Detection system for microcalcification cluster identification in a pan-European distributed database of mammograms}

\author{A. Retico\address[INFN]{Istituto Nazionale di Fisica Nucleare, Largo Pontecorvo 3, 56127 Pisa, Italy }\thanks{Corresponding author.  {\em E-mail address:} alessandra.retico@df.unipi.it (A. Retico).
{\em Tel:} +39 0502214459; {\em fax:} +39 0502214317.},
        P. Delogu\addressmark\address[DipPhys]{Dipartimento di Fisica dell'Universit\`a di Pisa, Largo Pontecorvo 3, 56127 Pisa, Italy},
	M.E. Fantacci\addressmark[INFN]\addressmark[DipPhys],
	A. Preite Martinez\addressmark[DipPhys],
 	A. Stefanini\addressmark[INFN]\addressmark[DipPhys]
and
        A. Tata\addressmark[INFN]}
       
\runtitle{CADe of microcalcification clusters in a pan-European database of mammograms}
\runauthor{A. Retico}

\begin{document}

\begin{abstract}

A computer-aided detection (CADe) system for microcalcification cluster identification in mammograms has been developed in the framework of the EU-founded MammoGrid project. 
The CADe software is mainly based on wavelet transforms and artificial neural networks. It is able to identify microcalcifications in different kinds of mammograms (i.e. acquired with different machines and settings, digitized with different pitch and bit depth or direct digital ones).
The CADe can be remotely run from GRID-connected acquisition and annotation stations, supporting clinicians from geographically distant locations in the interpretation of mammographic data.
We report the FROC analyses of the CADe system performances on three different dataset of mammograms, i.e. images of the CALMA INFN-founded database collected in the Italian National screening program, the MIAS database and the so-far collected MammoGrid images. The sensitivity values of 88\% at a rate of 2.15 false positive findings per image (FP/im), 88\% with 2.18 FP/im and 87\% with 5.7 FP/im have been obtained on the CALMA, MIAS and MammoGrid database respectively.

\vspace{1pc}
{\em Keywords}: Computer Aided Detection, Mammography, Wavelets, Neural Networks, GRID applications. 
\vspace{1pc}

\end{abstract}

\maketitle

\section{Introduction}

The EU-founded MammoGrid project~\cite{MammoGrid} is currently collecting an European-distributed database of mammograms with the aim of applying the  GRID technologies to support the early detection of breast cancer.
GRID is an emerging resource-sharing  model that provides 
a distributed infrastructure of interconnected 
computing  and storage elements~\cite{GRID}.
A GRID-based architecture  would allow the resource sharing and the 
co-working between radiologists throughout the European Union. 
In this framework, epidemiological studies, tele-education of young health-care professionals, advanced image analysis and tele-diagnostic support (with and without computer-aided detection) would be enabled.

In the image processing field, we have developed and implemented in a GRID-compliant acquisition and annotation station a  computer-aided detection (CADe) system able to identify microcalcifications in different kinds of mammograms (i.e. acquired with different machines and settings, digitized with different pitch and bit depth or direct digital ones). 

This paper is structured as follows: the detection scheme is illustrated in sec.~\ref{sec:Description}, sec.~\ref{sec:MammoGridDatabase} describes the database the MammoGrid Collaboration is collecting, whereas the tests carried out on different datasets of mammograms and the preliminary results obtained on a set of MammoGrid images are discussed in sec.~\ref{sec:TestRes}.

\section{Description of the CADe system}
\label{sec:Description}

The CADe procedure we realized is mainly based on wavelet transforms and artificial neural networks. Both these techniques have been successfully used in similar image processing applications~\cite{Richardson,Yoshida,Lado}. 
Our CADe system indicates one or more suspicious areas of a mammogram where microcalcification clusters are possibly located, 
according to the following schema:
\begin{itemize}
\item	INPUT: digital or digitized mammogram;
\item	Pre-processing: 
	a) identification of the breast skin line and segmentation of the breast region with respect to the background; b) application of the wavelet-based filter in order to enhance the microcalcifications;
\item	Feature extraction: a) decomposition of the breast region in several $N$$\times$$N$  pixel-wide partially-overlapping sub-images to be processed each at a time; % (N=60 pixels in the present analysis); 
 b) automatic extraction of the features characterizing each sub-image;
\item	Classification: assigning each processed sub-image either to the class of  microcalcification clusters or to that of normal tissue;
\item	OUTPUT: merging the contiguous or partially overlapping sub-images and visualization of the final output by drawing the contours of the suspicious areas on the original image.
\end{itemize}

\subsection{Pre-processing of the mammograms}

\begin{figure}[htb]
\vspace{9pt}
\begin{center}
\includegraphics[width=7.5cm]{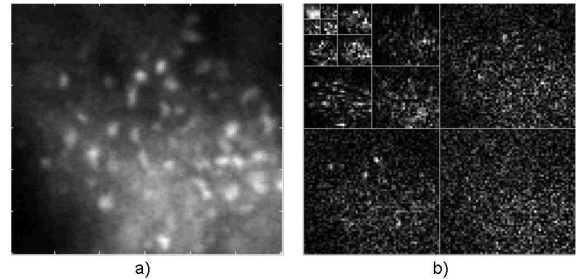}
\end{center}
\caption{Wavelet decomposition of a digitized mammogram ($85 \mu$m pitch): 
a) original image containing a
microcalcification cluster; b) 4-level
decomposition using Daubechies 5 mother wavelet.}
\label{fig:mammo2Dsub}
\end{figure}

The  pre-processing procedure aims to enhance the signals revealing the presence of  microcalcifications, while suppressing the 
complex and noisy non-pathological breast tissue.
A mammogram is usually dominated by the low-frequency information,
whereas the \mis appear as high-frequency 
contributions.
A particular of a mammographic image and its decomposition according to the
2-D multi-resolution analysis are illustrated in  fig.~\ref{fig:mammo2Dsub}.
It is worth noticing that \mis show some evident features at some
specific scales, while they are almost negligible at other scales.  
The use of the
wavelet transform allows for a separation of the more important
high-resolution components of the mammogram from the less important
low-resolution ones.

Once the breast skin line is identified, the breast region is processed by the  wavelet-based filter,  according to the 
 following main steps: identification of the
family of wavelets and the level up to which the decomposition has to be performed in order to highlight the interesting details; manipulation of the wavelet coefficients (i.e. suppression of the coefficients encoding the low-frequency contributions and enhancement of those  encoding the contributions of interesting details); inverse wavelet transform. 
By properly thresholding the wavelet
coefficients at each level of the decomposition, an enhancement of the
microcalcification with respect to surrounding normal tissue can 
be achieved in the synthesized image.
In order to achieve this result, the 
wavelet basis, the level up to which
the decomposition have to be performed and the thresholding rules
to be applied to the wavelet coefficients 
have to be accurately  set. All these choices and parameters 
are application dependent. The size of the pixel pitch and 
the dynamical range of the gray level intensities
characterizing the mammograms are the most important parameters to be
taken into account.

\subsection{Feature extraction}

In order to extract from a mammogram the features to 
be submitted to the classifier,  
small regions of a mammogram are analyzed each at a time. The choice of 
fragmenting the mammogram in small sub-images is finalized  
both to reduce the amount of data 
to be analyzed at the same time and to facilitate the localization of the 
lesions possibly present on a mammogram.
The size of the sub-images  has been chosen according to  
the basic rule of considering the smallest squared area matching 
the typical size of a small  \mi cluster.  
Being the size of a single \mi rarely greater  than 1 mm, and 
 the mean  distance between two microcalcifications belonging to the same 
cluster  generally smaller than 5 mm, 
we assume a square 
with a 5 mm side to be large enough to accommodate a small cluster. 
This sub-image size is appropriate  to 
discriminate an isolated microcalcification 
(which is not considered to be a pathological sign)
from a group of \mis close together.
The length of the square side in pixel units is obviously 
determined by the pixel pitch of the digitizer or of the direct digital device.
Let us assume that our choice for the length of the square side
corresponds to $N$ pixels.
In order to avoid the accidental missing of a \mi cluster happening to be at 
the interface between two contiguous  sub-images, we use the technique of the 
partially overlapping sub-images, i.e. we let the mask for 
selecting the sub-image to be analyzed
move through the mammogram by half of the side length 
($N/2$ pixels) at each horizontal and vertical step.
In this way each region of a mammogram is analyzed more than once  
with respect to different neighboring regions.

\begin{figure}[htb]
\vspace{9pt}
\begin{center}
\includegraphics[width=6.5cm]{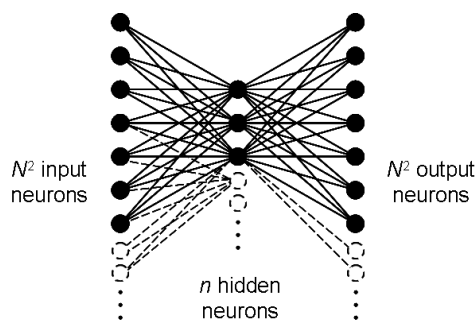}
\end{center}
\caption{Architecture of the auto-associative
neural network.}
\label{fig:autoassNN}
\end{figure}

Each  $N$$\times$$N$ pixel-wide sub-image extracted from the filtered mammogram
is processed by an
auto-associative neural network, 
used to perform an automatic
extraction of the relevant features of the sub-image.  
Implementing an auto-associative
neural network is a neural-based method for performing an unsupervised
feature extraction~\cite{Kramer1,Kramer2,Leonard,Kuespert}. This step has been introduced in the CAD
scheme  to reduce the
dimensionality of the amount of data (the gray level intensity values of
the $N$$\times$$N$ pixels of each sub-image) to be classified by the system.
The architecture of the network we use is a bottle-neck one (see fig.~\ref{fig:autoassNN}), consisting of three layers of
 $N^2$ input, $n$ hidden (where $n\ll N^2$)
and  $N^2$ output neurons respectively.
This neural network is trained to reproduce in output
the input values. The overall activation  of
the $n$ nodes of the bottle-neck layer summarize the relevant
features of the examined sub-image. The more the 
$N$$\times$$N$ pixel-wide sub-image  
obtained as output is close to 
the original sub-image provided as input,
the more the activation potentials of the $n$ hidden neurons 
are supposed 
to accommodate the information
contained in the original sub-image.

It is worth noticing that the implementation of an 
auto-associative neural network at this stage of the CAD scheme
allows for a strong compression of
the parameters representing each sub-image ($N^2 \to n$) 
to be passed to the following step of the
analysis.

\subsection{Classification}

We use the $n$ features extracted by the auto-associative neural
network 
to assign each sub-image   to either the class of sub-images 
containing microcalcification clusters
or the class of those consisting only of normal breast tissue.
A standard three-layer feed-forward neural network
has been chosen to perform the classification
of the $n$ features extracted from each sub-image. 
The general architecture characterizing  this net 
consists in  $n$  inputs, $h$ hidden and two output neurons, 
and the supervised training phase
is based on the back-propagation algorithm.

\section{The MammoGrid distributed database}
\label{sec:MammoGridDatabase}

One of the main goals of the EU-founded MammoGrid project is the realization of a GRID-enabled European database of mammogram, with the aim of supporting the collaboration among clinicians from different locations in the analysis of mammographic data.
Mammograms  in the DICOM~\cite{DICOM} format are collected
through the MammoGrid acquisition and annotation workstations installed in the participating hospitals. Standardized images are stored into the GRID-connected  database. The image standardization is realized by the Standard-Mammogram-Form (SMF) algorithm~\cite{SMF} developed by the Mirada Solutions Company$^{\rm TM}$, a partner of the MammoGrid project.
The SMF provides a normalized representation of the mammogram, i.e. independent of the data source and of the acquisition technical  parameters (as mAs, kVp and breast thickness).

\section{Tests and results}
\label{sec:TestRes}

As the amount of mammograms  collected at present in the MammoGrid database is too small for properly training the neural networks used in the characterization and classification procedures of our CADe, we used a larger dataset of mammograms for developing the system, then we evaluated its performances on the MammoGrid database.

The dataset used for training and testing the CADe was extracted from the fully annotated CALMA database~\cite{Bottigli,magic5} and it consists of 375 mammograms containing microcalcification clusters  and 610 normal mammograms, 
digitized with a pixel pitch of 85 $\mu$m and a dynamical range of 12 bit per pixel. 

\begin{figure}[htb]
\vspace{9pt}
\begin{center}
\includegraphics[width=7.5cm]{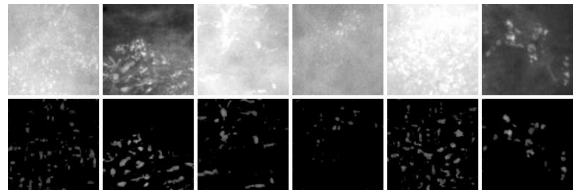}
\end{center}
\caption{ Examples of the wavelet-based filter  performances  (top/bottom: original/filtered sub-images containing microcalcification clusters).}
\label{fig:filtered-im}
\end{figure}

To perform the  multi-resolution analysis we considered the 
Daubechies family of wavelet~\cite{Daubechies}, in particular we used the
db$5$ mother wavelet.
As shown in fig.~\ref{fig:mammo2Dsub}, each sub-image is 
decomposed  up to the forth level. 
We found out that the
resolution level 1 mainly shows  the high-frequency noise included in
the mammogram, whereas the levels 2, 3 and 4 contain  the high-frequency
components related to the presence of microcalcifications.
Levels greater than 4 exhibited a strong correlation with larger structures 
possibly present in the normal breast tissue constituting the 
background.
In order to enhance microcalcifications, 
the approximation coefficients at level 4
and the detail coefficients at the first level 
were neglected. By contrast, the statistical analysis of the distributions of 
the remaining detail coefficients lead us to keep into account 
for the synthesis procedure
only those coefficients whose values exceed $2 \sigma$, where $\sigma$
is the standard deviation of the coefficient distribution at that level.
Some examples of the performance of the filter on mammographic images
containing
\mi clusters are shown in fig.~\ref{fig:filtered-im}. 

The training and testing of the auto-associative neural network has been performed on a dataset of 149 mammograms containing microcalcification clusters  and 299 normal mammograms. The best performances  were achieved with the following network architecture: 3600 input, 80 hidden and 3600 output neurons.
It corresponds to analyzing 60$\times$60 pixel-wide regions of mammograms each at a time. This size of the analyzing window is large enough to distinguish clusters from isolated microcalcifications since it approximately corresponds to a physical region of 5$\times$5 mm$^2$.
 
The dataset used for the supervised training of the feed-forward
neural classifier is constituted by 156 mammograms with 
\mi clusters  and 
241 normal mammograms. 
The standard back-propagation
algorithm was implemented and the best performance 
 were achieved with 10 neurons in the hidden layer.
%The average performances achieved in the testing phase 
%are  93.4\% for the sensitivity (true positive fraction) and 91.8\% for the specificity (true negative fraction). 
 
\begin{figure}[htb]
\vspace{9pt}
\begin{center}
\includegraphics[width=7.0cm]{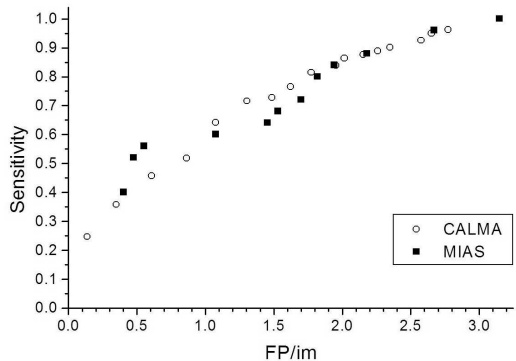}
\end{center}
\caption{FROC curve obtained on the CALMA dataset (140 mammograms) and on the MIAS   dataset (42 mammograms).}
\label{fig:FROC}
\end{figure}

The  CADe performances were globally evaluated on a test set of 140 images (70 with microcalcification clusters and 70 normal images) in terms of the free-response operating characteristic (FROC) analysis~\cite{Chakraborty} (see  fig.~\ref{fig:FROC}).  The FROC curve is obtained by plotting the sensitivity of the system versus 
the number of  false-positive detection per image (FP/im), 
while the decision threshold of the classifier is varied.
In particular, as shown in the figure, a sensitivity value of 88\% 
is obtained at a rate of 2.15 FP/im.

In order to test the generalization capability of the system, we evaluated the CADe performances on the public available MIAS database~\cite{Suckling}. Being the MIAS mammograms characterized by a different pixel pitch (50 $\mu$m instead of 85 $\mu$m) and a less deep dynamical range (8 bit per pixel instead of 12) with respect to the CALMA mammograms, we had to define a tuning procedure for adapting the CADe system to the database we used for this test. A scaling of the wavelet-analysis parameters (sum of four neighboring pixels, matching of the dynamical ranges, wavelet decomposition up to the third level) allows the CADe filter to generate very similar pre-processed images.
The remaining steps of the analysis, i.e. the characterization and the classification of the sub-images, have been directly imported from the CALMA CADe neural software. 
The performances the rescaled CADe achieves on the images of the MIAS database have been evaluated on a set of 42 mammograms (20 with \mi clusters and 22 normal) and shown in fig.~\ref{fig:FROC}.
As can be noticed, a  sensitivity value of  88\% is obtained at a rate of 2.18 FP/im.
The strong similarity in the trends of the FROC curves obtained on the CALMA and on the MIAS databases demonstrate the good generalization capability of the CADe system we developed.

\begin{figure}[htb]
\vspace{9pt}
\begin{center}
\includegraphics[width=7.0cm]{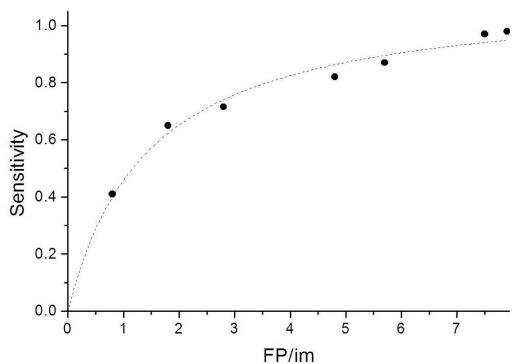}
\end{center}
\caption{FROC curve obtained on the  MammoGrid  database (130 mammograms).}
\label{fig:FROC_MG}
\end{figure}

According to the MammoGrid project work-flow~\cite{MammoGrid},  the CADe algorithm has to run on mammograms previously processed 
by the SMF software~\cite{SMF}.
The SMF mammograms  are characterized by a different pixel pitch (100 $\mu$m instead of 85 $\mu$m) and a different effective dynamical range (16 bit per pixel instead of 12) with respect to the CALMA mammograms.  The wavelet-analysis parameters have been rescaled  to run the CADe analysis on 
these images.
A test has been performed on a set of 130 mammograms with microcalcification clusters belonging to 57 patients: 46 of them have been collected and digitized at the University Hospital of Udine (IT), whereas the remaining 11 were acquired by the full-field digital mammography system GE Senographe 2000D at the Torino Hospital (IT); all have been stored in the MammoGrid database by means of the MammoGrid acquisition station installed at the University Hospital of Udine (IT).
The preliminary evaluation of the CADe performances in terms of the FROC analysis is shown in fig.~\ref{fig:FROC_MG}.
In this case a sensitivity value of  87\% is obtained at a rate of 5.7 FP/im.

\section{Conclusions}

We reported in this paper the details of the analysis and the results 
our CADe system for microcalcification cluster detection  achieves 
on mammograms collected in different hospitals and acquired with different 
methods. 
In particular, wavelet transforms have been implemented in the pre-processing step, in order to enhance the microcalcifications with respect to the complex and noisy patterns provided by the non-pathological breast tissue. This pre-processing method can be tuned 
on databases characterized by different pixel pitch or different dynamical range. The features to be used in the classification step are automatically extracted by means of an auto-associative neural network and then analyzed by a feed-forward neural network. 

The CADe system we developed and tested on both the CALMA and the MIAS databases has been adapted to the MammoGrid SMF images by re-scaling the wavelet-filter parameters. 
The main advantage the scaling procedure provides is that the rescaled CADe can be run even on small databases not allowing for a proper training of a neural decision-making system to be carried out. 
The preliminary results obtained on the so-far collected MammoGrid database are encouraging. Once the planned increase in the population of the database is realized, a complete  test of the CADe performance on the pan-European MammoGrid database would be carried out.

\end{document}